\documentstyle[preprint,aps,12pt]{revtex}
\newcommand{\la}{\langle}
\newcommand{\ra}{\rangle}

\newcommand{\beq}{\begin{eqnarray}}
\newcommand{\eeq}{\end{eqnarray}}

\renewcommand{\Im}{{\rm Im}}

\newcommand{\btem}{\bibitem}
\newcommand{\Pai}{{\rm \Pi}}
\newcommand{\non}{\nonumber}
\newcommand{\GeV}{{\rm GeV}}
\tighten
\begin{document}
\preprint{UTHEP-329, January 1996}
\draft
\flushbottom

\title{  Second Class Current in QCD Sum Rules}

\author{H. Shiomi\footnote{e-mail address: siomi@nucl.ph.tsukuba.ac.jp}}

\address{ Institute of Physics, University
 of Tsukuba, Tsukuba, Ibaraki 305, Japan}

\maketitle
 
\centerline{(\today)}

\vspace{1.5cm}

\begin{abstract}
 Induced tensor  charge of the nucleon $g_T$,
 which originates from  G-parity violation, 
 is evaluated from QCD sum rules.
We find that  $g_T/g_A$ with $g_A$ being the axial charge 
is $ -$0.0152 $\pm$ 0.0053 which is proportional to u-d quark
 mass difference.
This result is small compared to  preliminary analysis of
 the experiment, but is consistent with the estimate in the MIT
 bag model.
\end{abstract}

\vspace{0.4cm}

%
\section{Introduction}

 Deriving the coupling constants from QCD \cite{Muta87} is  one
of the most important themes in hadron physics. 
Also,  getting precise values of
 the coupling constants from the first principle   
 will enable us to make more  quantitative  predictions for nuclear systems. 
        
For extracting the hadronic quantities
  from QCD,   QCD Sum Rules (QSR) discussed below
 are  known to be one of the two powerful tools. (The other 
 is lattice QCD simulations \cite{Lattice}.)
 QSR  was first proposed in a paper by Shifman Vainsthein,
 and Zakharov in 1978 \cite{SVZ78}, in which the main idea and 
the application to meson systems such as
meson masses, decay rates and $\rho-\omega$ mixing, are shown.     
Later, many applications have been made \cite{Shf92,RRY85}.
The extension  to  the baryon systems
  was put forward by Ioffe \cite{Ioffe81}.
 QSR with external field  was  also proposed by  Ioffe and  Smilga
 \cite{IS84}, which will be discussed later.

In this paper we will evaluate induced tensor charge of the nucleon 
$(g_T)$ with the help
 of axial charge $(g_A)$ and nucleon sum rules. 

 $g_T$ and $g_A$ are defined by the
 nucleon matrix element of the axial current,
 \beq
\label{axial-mat}
& &\la P(p_2)| A_\mu^5 | N(p_1) \ra   \nonumber \\
& &
=\bar{u}_p(p_2) (\gamma_\mu \gamma_5 g_A 
+ { \gamma_5 g_P \over M_p +M_n}
+{ i \gamma_5 \sigma_{\mu \nu} q^{\nu}
 \over M_p+M_n} g_T  ) u_n(p_1), \ \ \ q=p_1-p_2,
\eeq
where $M_p(M_n)$ is proton (neutron) mass, $A_\mu^5=\bar{u}
 \gamma_\mu \gamma_5 d$
 represents axial  current
and $u_p (u_n)$ reveals proton (neutron) wave function.
  $g_P$ is called induced psudoscalar constant, which will not
 be  examined
in this paper.
The deviation of $g_A$ from unity is a reflection of
 the underlying composite  structure of the nucleon and  
there have been many  studies  on $g_A$, e.g., QCD sum rules \cite
{BK84,pasu}, quark models 
\cite{Kok69} , the bag model \cite{DeG75} and the skyrme model \cite{ANE83}. 
 The difference of  $g_A$ and $g_T$ is classified by 
$G$-parity which is the charge 
conjugation $C$ 
combined with the  rotation of $180^\circ$  around the y-axis in the isospin space 
\beq
G=C e^{i I_y \pi},
\eeq
where $I_y$ is the rotational matrix around the y-axis in the isospin space.

Under the $G$-parity, 
\beq
\label{G-trans}
G \bar{p} \gamma_\mu \gamma_5 n G ^{-1}
=- \bar{p} \gamma_\mu \gamma_5 n,  \ \ \
G \bar{p} \sigma_{\mu \nu} \gamma_5 n G^{-1}
= \bar{p} \sigma_{\mu \mu} \gamma_5 n, \ \ 
G A^5_\mu G^{-1}= -A^5_\mu .
\eeq
The first current in eq.(\ref{G-trans}) with the same sign as that of 
$A_\mu^5$ under $G$- parity
 is called  the {\it first class current}, while the second current
 in eq.(\ref{G-trans}) with the opposite sign as that of $A^5_\mu$
 is referred  to as  the {\it second class current} \cite{Wein58}.       

There are two sources of $G$-parity violation in the standard model.
 One is from QED (electric charges of $u$ and $d$ quarks are
 different) and another is from the
 mass term in the  QCD Hamiltonian 
 (masses of $u$ and $d$ quarks are different).
 We will exclusively examine the latter effect in this thesis.

 The mass term in the QCD Hamiltonian is written as
\beq
\label{QCDmass}
H_{mass}={1 \over2 }(m_u+m_d)(\bar{u}u+ \bar{d} d)
 +{1 \over2 }(m_u-m_d)(\bar{u}u- \bar{d} d).
\eeq
where
the light quark masses 
 are determined from analyses of  the hadron mass splittings in 
 QCD sum rules \cite{Narison87};
\beq
& &m_u(\mu=1{\rm GeV}) \ = \ (5.1 \pm 0.9){\rm MeV},
 \ m_d(\mu={\rm 1GeV})=(9.0 \pm 1.6) {\rm MeV}.
 \nonumber
\eeq

Under G-parity,
the first term  in eq.(\ref{QCDmass}) does not change
the sign,  but the second term in eq.(\ref{QCDmass})
 changes sign, which means that
 $g_T $ is  represented as $g_T \sim (m_u- m_d)/M_N$ since
 $g_T$ is  dimensionless. 
 This implies that   $g_T$ will be much smaller than $g_A$ because of 
the small $u-d $ quark mass difference
 $g_T \sim 4 {\rm MeV}/1000 {\rm MeV} \sim 0.004$ where we have used
 the quark mass difference $m_d-m_u \sim 5 {\rm MeV}$.

This rough estimate of  $g_T$ using  
  the $G$-parity violation, however, may not be consistent 
 with the analysises of the experimental data given by measuring the 
 beta-ray angular distribution in
aligned ${}^{12}{\rm B }$ and ${}^{12} {\rm N}$ \cite{Morita9}.
In Ref.\cite{Morita9},   results of the analyses
 using the experimental data  are quoted as 
\beq
& & g_T / g_A=0.14 \pm 0.10 \ \ \mbox{in 1985}, \\
& &  g_T/ g_A=-0.21 \pm 0.14 \ \ \mbox{in 1992}. 
\eeq
This shows that $g_T$ is of order 10 $\%$ compared to
$g_A$, which is  order of magnitude larger than the naive expectation.
Although the experimental error bars are large and even the sign of $g_T/g_A$
is not certain yet, the data poses a theoretical challenge to give more reliable
estimate of $g_T/g_A$. 

It is in order here to show
 other examples of the G-parity violation which is 
 more firmly established than $g_T$ \cite{Hatsu94}:\\
1)Proton-neutron mass difference.\\
 Experimental mass difference is $M_p-M_n=-1.29$ MeV, 
in which the contribution of
 the $u-d$ quark mass difference after  subtracting the theoritical
 electromagnetic effect (0.76$\pm$0.3 MeV)
is $-$2.05 MeV.  This last number has been successfully reproduced
 in QSR calculations \cite{HHP90}.\\ 
2)$\rho^0-\omega$ mixing. \\ 
The $\rho- \omega$ mixing is defined by the covariant matix element
 $\la \rho^0| H_{GPB}| \omega \ra$ at the $\rho^0-\omega$ mass shell
 with $H_{GPB}$ being the second term in eq.(\ref{QCDmass}).
The recent measurement of the $e^+ e^- \rightarrow \pi^+ \pi^-$ shows
 an unambigous determination of the $\rho^0-\omega$ mixing with negative sign,
 which ought to be  dominated by the quark mass difference since
  the electromagnetic effect by $\rho \rightarrow \gamma \rightarrow \omega$ 
  is positive and small\cite{SVZ78a,HHMK94}.\\
3)$\pi^\pm- \pi^0$ mass differnce ($m_{\pi^\pm}-m_{\pi^0}$=4.6 MeV).\\  
This is a typical example of the electromagnetic
 $G$-parity violation.
 Theoretical estimate gives 
 $(m_{\pi^\pm}-m_{\pi^0})_{em}=4.6\pm0.1 $ MeV, while
 the effect of the quark mass difference appears only in second order
 of the quark masses and is extremely small\cite{BBG89}.

  The ingredients  of this paper  are twofolds: i) to get 
 QSR with an external field for $g_T$ and $g_A$ in section II-V,
 and  ii) to predict the value of $g_T$ relative to $ g_A$ in section VI.

As for i), we will adopt a method
 proposed by loffe and Smilga \cite{IS84} and independently
 by Balitsky and  Yung \cite{BY83}, in which two point functions with  
 an external electromagnetic field strength  $F_{\mu \nu} $ is 
 studied up to linear in $F_{\mu \nu}$.
 The method
  has been applied for the magnetic moment of the nucleon and 
 the results agree with the experimental data with a good  accuracy.
 A method on two point functions with an 
  axial-vector field $ Z_\mu$ 
   was also developed by Belyaev and Kogan \cite{BK84}, and later 
improved by
Pasupathy et al. \cite{pasu}.
 They have considered terms proportional to $Z_\mu$  for
 evaluating the axial charge $g_A$.
 The latter method with $Z_{\mu}$ replaced by the vector
 potential $A_{\mu}$ is, however,  not suitable
 for studing  magnetic  moments
 since the explicit  momentum transfer 
 must be retained.
 
 Adopting   QSR with the external field  induces
  new parameters which are absent
  in  ordinary QSR. These parameters
   reflects the response of QCD
  vacuum to  the external field.
 For instance, $\la 0| \bar{q} \sigma_{\mu \nu} q |0 \ra_E$, 
 which is identical to zero 
 in the vacuum,
  acquires the non-zero value due to the presence
 of the external field.   
 To evaluate these new condensates, QSR with the
 assumption of the vector dominance can be used
  \cite{BK}. 

 There is another new feature of QSR with the external field
 compared to the ordinary one. 
The phenomenological side of the correlation
 function with external field takes the following double pole
 form near the nucleon resonance 
\beq
\la 0 | \eta | N \ra 
 \la  N | J^E | N \ra \la N |\bar{\eta}|0 \ra (p^2 -M_N^2)^{-2},
\eeq
where $J^E$ is  a current coupled to the external field.
 Besides the double pole part which we are interested in,
  single poles, which expresses transition from
 the ground state to  excited states, appears.
 Furthremore, the single pole term is not suppressed compared
 to the double pole term  after 
 applying the Borel transform.
 This bears no resemblance to the contribution of  continuum 
 which is exponentially suppressed by Borel transform.    
 Hence  we must take into account  both the double pole and the single pole
 in phenomenological side on the same footing, which  
requires   a  procedure to subtract the single poles. 

 As for ii), one must remember that   
 $g_T$  originates from 
 the  $G$-parity violation induced by  the u-d quark mass difference and
 the electromagnetism. The experimental value of $g_T$ is still 
 uncertain as we have mentioned above.
 Thereby we shall try to determine $g_T$ from QSR with
  main emphasis on the effect of the u-d quark mass difference.
 Within our knowledge, no serious evaluation of 
 $g_T$ has been done so far except for a rough
 estimate using the MIT bag model \cite{DH82,Hol89}.
 We will therefore reexamine the bag model calculation also
 and compare it with our QSR result.

The paper is organized as follows.
Section II-IV are devoted to derive $g_T$ and $g_A$ sum rules
 in QSR with the external field.
In section V, we estimate the quark and induced  condensates by
 using QSR.
In section VI, we analyse the $g_T$ sum rule and get its numerical number.
 In section VII, discussions  and summary are made. 
\\ 

\section{ Weak interaction in  hadronic side and QCD side }
We start with the  two point function with an external field;
\beq
\label{corre-ext}
\Pai_E(p)&=&i \int d^4x e^{ip \cdot x} \la 0|T{\eta_p(x)
 \bar{\eta}_n(0) }|0 \ra_E \\
     &=& F_{\mu \nu}\Pai_{\mu \nu}(p) ,
\eeq
where 'E' denotes  external field of weak boson $W^+$,
 $F_{\mu \nu}(x)=\partial_\mu W^+_\nu (x)-\partial_\nu W^+_\mu (x)$, and
 $\eta_p (\eta_n)$ corresponds to
the proton (neutron) interpolating field  defined as \cite{Ioffe81}
\beq
\eta_p(x)
= \epsilon_{abc} (u^a(x) C \gamma_\mu u^b(x)) \gamma_5 \gamma_\mu d^c(x),
 \ \ \ \eta_n(x)=\eta_p(u \leftrightarrow d)
\eeq
 The hadronic side of sum rules can be saturated by a 
 process where
  neutron turns into  proton by absorbing the $W^+$ boson, matrix element
 of which is shown in eq.(\ref{axial-mat}).
 Hence we define  an  effective Lagrangian in which nucleon current couples
  to  $W^+$ field as
\beq
\label{int-had}
L^{had}_{int}=-{g \over 2 \sqrt2 }j_\mu^5 W^+_\mu
=-{g \over 2 \sqrt{2}}\bar{p}
\left ( g_A \gamma_\mu \gamma_5  W^+_\mu + 
{g_T \over M_p+ M_n} \gamma_5
 \sigma_{\mu \nu} \partial_\nu W^+_\mu \right)n  ,
\eeq   
where p (n) represents  proton (neutron) field
and $g$ is associated with
 Fermi constant as ${G_F \over \sqrt{2}} ={g^2 \over 8 M_W^2}$
with $M_W$ being $W^+$ boson mass 
 in the Glashow-Weinberg-Salam model \cite{Wein67}.

It is in order here to mention  $g_p$.
 $g_P$ is associated with $g_A$ via PCAC and can be directly measured 
by  the muon capture \cite{Hol89}. 
 However, our external field 
 does not pick up the contribution of  $g_P$
because  we are using the field strength $F_{\mu \nu}$
 instead of the vector potential $W_{\mu}^+$ as an external field.

On the other hand, in the quark level, 
 the interaction of quarks and  the external field
 is written as 
\beq
\label{int-QCD}
L_{int}^{quark}={-g \over 2 \sqrt2}j_\mu^5 W^+=
{-g \over 2 \sqrt{2}} \bar{u} \gamma_\mu \gamma_5 d \ \ W_\mu^+ ,
\eeq
where $u$ and $d $ are up and 
  down quark respectively.
   The common factor
${g / {2 \sqrt2}}$ in eq.(\ref{int-had}) and 
 in eq.(\ref{int-QCD})
  is obtained by   comparing  the  V-A theory \cite{GM57}
 with the Glashow-Weinberg-Salam model.\\
%
\section{ Hadronic side for two point function with external field}
In this section we examine the hadronic   contribution
 to  QSR with the external field.
Firstly,  
 we consider  Fig.1, which  shows  that
  a neutron with momentum $p_1$ absorbs
  $ W^+$ boson and turns into proton with momentum $p_2$.
Using eq.(\ref{int-had}), we may write down Fig.1 as   
\beq
\mbox{Fig.1}=& &{ 1 \over \hat{p}_2 -M_p} 
\left ( g_A \gamma_\mu \gamma_5
+{ i \gamma_5 \sigma_{\mu \nu} \over M_p +M_n} g_T q_\nu \right )
{ 1 \over \hat{p}_1 -M_n} \\
& &
={1 \over (p^2_2- M_p^2)(p_1^2-M_n^2)} \nonumber  \\
& &
\label{pheno-sub}
\times \left (-{\hat{q} \over 2}+\hat{p} +M_p \right) 
\left ( g_A \gamma_\mu \gamma_5 
+{ i \gamma_5 \sigma_{\mu \nu} \over M_p +M_n} g_T q_\nu \right )
\left ({\hat{q}  \over 2}+\hat{p} +M_n \right) , 
\eeq   
where $p_2 (p_1)$ is proton (neutron) momentum mentioned above,
 $p={p_1 +p_2 \over 2}$ and $q=p_1-p_2$.
 In the limit of soft external momentum
  $q_\mu \rightarrow 0$,
  we keep only the terms proportional to $q_\mu$  in eq.(\ref{pheno-sub})
to extract   $F_{\mu \nu}$.
Then eq.(\ref{pheno-sub}) is reduced to 
\beq
{ i \gamma_5 q_\nu  \over (p^2- M_n^2)(p^2- M_p^2) }
\left [ P_1 \hat{p} \sigma_{\mu \nu}
+ P_2\sigma_{\mu \nu} \hat{p} 
+ P_3\sigma_{\mu \nu} +P_4(\gamma_\mu p_\nu- \gamma_\nu p_\mu)
 \hat{p} \right] 
\equiv
q_\nu  \Gamma_{\mu \nu}(p) ,
\eeq
where
\beq
\label{pheno-p1}        
& & P_1=-{g_A \over 2}-{M_n \over M_n +M_p}g_T , \\
& &
P_2=
-{g_A \over 2} +{M_p \over M_n + M_p} g_T , \\
& &
P_3=
-{g_A \over 2}(M_n- M_p) +({M_n M_p- p^2 \over M_p + M_n})g_T ,  \\
& &
\label{pheno-p4}
P_4=
{2 i g_T \over M_p + M_n } .
\eeq
By using eq.(\ref{int-had}), we evaluate  eq.(\ref{corre-ext}) 
 in first order of the external field:
\beq
& & -\int d^4 xe^{i p \cdot x}
 \la 0| T(\eta_p(x) \bar{\eta}_n(0) L_{int}^{had}) | 0 \ra \nonumber \\
 &=& -{g \over 2 \sqrt{2}} \lambda_p \lambda_n \int d^4 y {1 \over (2 \pi)^4}
\int d^4 l e^{i(p-l)\cdot y } W^+ (y) \nonumber \\
& & \times {1 \over \hat{p} -M_p}
\left ( g_A \gamma_\mu \gamma_5 
+{ i \gamma_5 \sigma_{\mu \nu} \over M_p +M_n} g_T (l-p)_\nu \right )
{ 1 \over \hat{l} -M_n}   \nonumber \\
 &=& -{g \over 2 \sqrt{2}} \lambda_p \lambda_n  {1 \over (2 \pi)^4}
\int d^4 y \int d^4q e^{-iq\cdot y } q^\nu \Gamma_{\mu \nu}(p) W^+ (y)  \nonumber \\
&= &-{i g \over 4\sqrt{2}}
 \lambda_p \lambda_n \Gamma_{\mu \nu} F_{\mu \nu}(0)  \nonumber \\
&= &{g \gamma_5 \lambda_n \lambda_p \over 4 \sqrt{2} (p^2-M^2_n)(p^2-M^2_p) }
 F_{\mu \nu}(0)  \nonumber \\
& &
\label{pheno-final}
\times \left [ P_1 \  \hat{p} \sigma_{\mu \nu}
+ P_2  \ \sigma_{\mu \nu} \hat{p} + P_3 \ \sigma_{\mu \nu} +P_4 \
(\gamma_\mu p_\nu- \gamma_\nu p_\mu)
 \hat{p} \right]  ,
\eeq
where $P_1, P_2, P_3$ and $P_4$
 are defined  in eq.(\ref{pheno-p1})-eq.(\ref{pheno-p4}),
 $ \lambda_n $ and $\lambda_p$ 
 are defined as $\la 0 |\eta |N \ra= \lambda_N u(p)$ with
$u(p)$ being the nucleon Dirac spinor.
Apart from  the  terms above, we must take into account 
 two other contributions.
One is  the single pole caused by a transition of
 nucleon to resonance states as follows,
\beq
\Pai_E(p)\sim \lambda_N \lambda_{N^*}
 {1 \over \hat{p}-M_N} H_{N N^*}{1 \over \hat{p}-M_{N^*}} ,
\eeq
where $N (N^*)$ is the nucleon (excited states, e.g., $N$(1440)), and
$H_{N N^*}$ is a  transition matrix from the nucleon to the 
 excited states. As we have mentioned, 
 the single pole 
 is not suppressed compared to the 
 double pole after  the Borel transform.
  Since we are not interested in the single poles,
 we will subtract them,  using a procedure shown later.
The other hadronic contribution 
 is  a continuum starting
  at a threshold  $S_0$,  which contains only 
 the excited states.
 The continuum can be  exponentially
 suppressed by applying the Borel transform. \\   
\section{ OPE for two point function with external field}
In ordinary QSR with no external fields,
 only Lorenz invariant operators  survive in OPE. 
 On the other hand, 
the external field induces new condensates. 
 Relevant  condensates up to dimension 6 in our case read  
\beq
\label{cond1}
 \la 0|\bar{u} \gamma_5 \sigma_{\mu \nu} d |0 \ra_E &=&
(m_u-m_d)F_{\mu \nu} {g \over 2 \sqrt2} \chi (0)  ,  \\
\label{cond2}
 g_s \la 0|\bar{u} \gamma_5 G_{\mu \nu} d |0 \ra_E &=&
\la \bar{d}d-\bar{u}u \ra_0 F_{\mu \nu}{g \over 2\sqrt2} \kappa (0)  , \\
 g_s  \epsilon_{\mu \nu \rho \omega} \la0|\bar{d} G_{\rho \omega} u|0\ra_E  ,
&=& 
\label{cond3}
\la \bar{d}d-\bar{u}u \ra_0 i F_{\mu \nu}{g \over 2\sqrt2} \xi(0) .
\eeq
 The above condensates are non-vanishing because 
 the QCD vacuum is distorted by the external field and the
Lorenz invariance is broken.
 Note also that  taking  $m_u= m_d$ makes
all the above terms vanish.
Using the fixed point gauge, we can rewrite 
 the $i \la 0| T(\eta_p(x) \bar{\eta}_n(0)) |0 \ra_E$ as follows:
\beq
& & i \la 0|\ T(\eta_p(x) \bar{\eta}_n(0))\ |0 \ra_E=
4 i \epsilon_{abc} \epsilon_{def}
\la 0|\ \gamma_5 \gamma_\mu i S_d^{ce}(x) \gamma_\nu C
i E_{ad}^T(x) C \gamma_\mu i S_u^{bf} \gamma_\nu \gamma_5 \ |0 \ra_E  \non \\
 & & \label{E-OPE} \\
\mbox{with}& & \non
 \\
& & iS_q^{ab}(x)=\la 0|T(q^a(x) \bar{q}^b(0))|0\ra \nonumber \\
& &
\label{Spropa}
 ={i\hat{x} \over 2 \pi^2 x^4} \delta^{ab}
+ {ix_\alpha  \over 8\pi^2 x^2}
(t^e)^{ab} \tilde{G}^{\alpha \rho}_e \gamma_\rho \gamma_5
-{m \delta^{ab} \over 4 \pi^2 x^2}   
+\la \chi^a_q (x) \bar{\chi}^b_q (0) \ra_0  , \\
& & iE^{ab}(x)=\la 0|T(u^a(x) \bar{d}^b(0)|0 \ra_E  \nonumber \\
& &
\label{Epropa}
={ig \over 2 \sqrt2} \delta^{ab}
{x_\lambda \over 8\pi^2 x^2} \tilde{F}_{\lambda l} \gamma_l
 + \la \chi^a_u(x) \bar{\chi}^b_d (0) \ra_E 
 \\
& &
-(m_u -m_d)\left \{{1 \over 32\pi^2}(\log(-x^2 \Lambda^2/4)+2 \gamma_E)
\gamma_5 \sigma_{\rho \nu} F^{\rho \nu}
+{ i \over 16 \pi^6 x^2}
 F_{\rho \mu} \gamma_5 \gamma_\mu \hat{x}x_\rho \right \} \non  ,
\eeq
where $i E^{ab}(x) $ is calculated by eq.(\ref{int-QCD}) with the first order
perturbation, $\tilde{F}_{\mu \nu}={1 \over 2}\epsilon_{\mu \nu \rho \omega}
F^{\rho \omega}$,  and $\Lambda^2$ is an infrared cut off parameter.

 $\la \chi^a_u(x) \bar{\chi}^b_d(0) \ra_E $ in eq.(\ref{Epropa})
 expresses non-perturbative condensate under the
external field, while other terms
  are coupled directly to the quark propagator.   
 To calculate  the nonperturbative terms, we shall
compare it with  $\la \chi^a_q(x) \bar{\chi}_q(0) \ra_0$ which
 is expanded as 
\beq
\la \chi^a(x) \bar{\chi}^b(0) \ra_0 &=&\la 0| q^a(x) \bar{q}^b(0) |0 \ra \\    
&= &-{\delta^{ab} \over 12} \la 0| \bar{q} q|0\ra - {\delta^{ab} x^2 \over 192}
\la 0| g_s \bar{q} \sigma \cdot G q |0 \ra  +  \cdots
\eeq
In the case above, we have retained  only the Lorentz scalar operators.
 In contrast, $\la\chi^a_u (x) \chi^b_d (0) \ra_E$ is expanded  only by  
 the  Lorenz tensor terms
 corresponding to  
 the induced condensates eq.(\ref{cond1})$-$ eq.(\ref{cond3}).
Hence 
\beq
\la \chi^a_u(x) \bar{\chi}^b_d(0) \ra_E
 &=&-{1 \over 24} \delta_{ab} \gamma_5\sigma_{\mu \nu}
 \la \bar{u} \gamma_5 \sigma_{\mu \nu} d \ra_E
- {x_\rho x_\omega \over 48} \gamma_5 \sigma_{\mu \nu} 
\la \bar{u} \gamma_5 \sigma_{\mu \nu}
 D_\rho D_\omega d \ra_E  + \cdots \nonumber \\
&= &
-{1 \over 24}{g \over 2 \sqrt2} 
\delta^{ab} \gamma_5 \sigma_{\mu \nu}F^{\mu \nu}
(m_u-m_d) \chi(0) \nonumber \\
& &-{g \over 2 \sqrt2}{1 \over 3^2 2^5} \gamma_5 \sigma_{\mu \nu}
\la \bar{d} d- \bar{u} u \ra_0  \nonumber \\
& & \times \left  \{(\kappa (0)-\xi (0) ) x^2 F_{\mu \nu}
-(2 \kappa (0)+ \xi (0) ) x_\mu x_\omega F_{\nu \omega} \right \} + \cdots
\eeq

We turn to carry out  OPE for eq.(\ref{E-OPE}) with diagrams Fig. 2(a)-(j).
For the  chiral odd structures,  we impose $m_u=m_d$.
This induces $P_1=P_2=-g_A/2 $ in eq.(\ref{pheno-final}) due to $g_T=0$.
 The chiral odd structure can be  applied to  
  estimate  the axial-charge ($g_A$) with  $m_u =m_d $.

On the other hand, 
 the chiral even structures in eq.(\ref{E-OPE}) are evaluated
 up to linear in $(m_u-m_d)$, which leads to $g_T$ sum rule. 

 Let us examine each contribution in Fig.2 more closely\\
The coefficient of $F_{\mu \nu}$ with
 the chiral odd structure given in Fig.2(a) reads    
\beq
\mbox{Fig.2(a)} &=&- {g \over 2 \sqrt2}{6 \over \pi^6 x^8}x_\lambda \gamma_l
\tilde{F}_{\lambda l}  . 
\eeq
The coefficient of $F_{\mu \nu} \la {\alpha_s \over \pi}G^2 \ra$ 
with the chiral odd structure
 given in Fig.12(b) reads
\beq
\mbox{Fig.2(b)}=-{g \over 2\sqrt2}
{1 \over 32 \pi^4 x^4} x_\lambda \tilde{F}_{\lambda l} \gamma_l
\left \la {\alpha_s \over \pi  } G^2 \right \ra_0 ,
\eeq
where $\left \la {\alpha_s \over \pi  } G^2 \right \ra_0 = 0.012 (\GeV^2)$
 is called gluon condensate, whose value is determined from the analysis
of heavy quark system based on QSR.  
The coefficient of $F_{\mu \nu} \la \bar{q}q\ra^2$ 
with the chiral odd structure given  in Fig.12(c) reads 
\beq
\mbox{Fig.2(c)}&=&-{g \over 2\sqrt2}{ \la \bar{q}q \ra^2_0  \over
12 \pi^2 x^2} x_\lambda
 \tilde{F}_{\lambda l} \gamma_l  . 
\eeq
The coefficient of $(m_u-m_d)F_{\mu \nu}$ coupled directly to the propagator
 given in Fig.2(d) reads
\beq
 \mbox{Fig.2(d)}=-i{g \over 2\sqrt2}{3 \over 2 \pi^6 x^8} x_\lambda
\tilde{F_{\lambda \l}} \gamma_l \hat{x}(m_u-m_d)  . 
\eeq
Fig.2(f) also gives a
 coefficient of $(m_u-m_d)F_{\mu \nu}$. 
 However, it causes the infrared divergence which 
 must be absorbed into the same dimensional  operator shown in Fig.2(e).\\ 
The coefficient of
 $\la \bar{d}\gamma_5 \sigma_{\mu \nu}u \ra_E$, which is  dimension 3
  and is given in Fig.2(e), reads
\beq
\mbox{Fig.2(e)}=
i {1 \over \pi^4 x^8} \gamma_5 \hat{x} \sigma_{\rho \omega}\hat{x}
\la \bar{d} \gamma_5 \sigma_{\rho \omega} u\ra_E   . 
\eeq
The coefficient of $\la \bar{d}d-\bar{u}u \ra_0 F_{\mu \nu}$
 given in Fig.2(g) reads
\beq
\mbox{Fig.2(g)}&=&-i{g \over 2 \sqrt2} {1 \over 2 \pi^4 x^6}
x_\lambda \tilde{F}_{\lambda l} \hat{x} \gamma_l
\la \bar{d} d-\bar{u} u \ra_0  . 
\eeq
The coefficient of  $ g_s \la \bar{d}\gamma_5 G_{\mu \nu} u \ra_E
$ and  $ g_s \epsilon^{\mu \nu \rho \omega}\la \bar{d}
 G_{\rho \omega} u \ra_E$ given in Fig.12(h) and (i) read
\beq 
\mbox{Fig.2(h) and (i)}&=&i{\gamma_5 
\sigma_{\rho \omega} \over 12 \pi^4 x^4}
[ g_s \la  \bar{d} \gamma_5 G^{\rho \omega} u \ra_E
+i g_s \epsilon^{\rho \omega \mu \nu} \la \bar{d} G_{\mu \nu} u \ra_E ] \\
& +&
{\gamma_5(\gamma_\rho x_\omega-\gamma_\omega x_\rho) \hat{x}
 \over 4 \pi^4 x^6}[g_s \la \bar{d} \gamma_5 G^{\rho \omega} u \ra_E
 +{1 \over 2}g_s i \epsilon^{\rho \omega \mu \nu} \la \bar{d}G_{\mu \nu} u \ra_E]  , 
\eeq
where the contribution of Fig.2(h) is zero bacause $\la 0|\bar{d}u |0\ra=0$
  and in Fig.2(i) the gluon field emitted by a soft quark
 interacts $W^+$ field via quark condensate. \\
The coefficient of  $ g_s \la \bar{d}\gamma_5 G_{\mu \nu} u\ra_E$ and  
 $ g_s \epsilon^{\mu \nu \rho \omega}\la \bar{d} G_{\rho \omega} u \ra_E$
 given in Fig.2(j) read
\beq
\mbox{Fig.2(j)}&=&
-i{\gamma_5 (\gamma_\rho x_\omega-\gamma_\omega x_\rho) \hat{x}
 \over 8 \pi^4 x^4} \hat{x}
[g_s\epsilon^{\rho \omega \mu \nu} \la \bar{d} G_{\mu \nu} u \ra_E
-2 i g_s \la \bar{d} \gamma_5 G^{\rho \omega} u \ra_E] \\
& &
-i {\gamma_5 \sigma_{\rho \omega} \over 4 \pi^4 x^4}
g_s \la \bar{d} \gamma_5 G^{\rho \omega} u\ra_E  ,
\eeq
where the gluon field  emitted by a hard quark interacts $W^+$ field
via quark condensate.

 In summary, we obtain the following formula  
\beq
\Pai_E(p)& =&{g \over 2 \sqrt2}F_{\mu \nu} 
\left[ Q_1 \gamma_5 (\hat{p} \sigma_{\mu \nu}+ \sigma_{\mu \nu} \hat{p})+
\{Q_2 \gamma_5 \sigma_{\mu \nu} + Q_3 i \gamma_5 
(\gamma_\mu p_\nu-\gamma_\nu p_\nu)
\hat{p}\} (m_u-m_d) \right]  , \non \\
& &  
\label{ope-final}
\\
Q_1&=&{-1 \over 32 \pi^4} p^2 \log(-p^2)-
{1 \over 64 \pi^2}{{\la {\alpha_s \over \pi}
G^2 \ra} \over p^2}
-{\la \bar{q} q\ra^2_0  \over 6 p^4} , \\
Q_2 &=& -p^2 \log (-p^2) \left({1 \over 16 \pi^4}-{\chi(0) \over 24 \pi^2}
 \right)-
C_m 
\left({ 1\over 8\pi^2} -{\kappa (0) \over 6 \pi^2}-
{\xi(0) \over 12 \pi^2} \right) \log(-p^2) ,  \\
Q_3&= &\log (-p^2) \left ({1 \over 32 \pi^4}- {\chi(0)  \over 12 \pi^2} \right)
+{C_m \over 8 \pi^2 p^2}   ,
\eeq
where we have used  a relation $\la \bar{d}d-\bar{u}u\ra_0 =C_m(m_u-m_d)$ 
 which will be discussed below.
 \\
\section{ The estimate of the quark and induced condensates}
Before setting up  the sum rules, we need to estimate
 the magnitude of  the quark and induced condensates.
For the quark condensate, 
 we utilize Finite Energy Sum Rules (FESR) \cite{KPTS83,SH95} 
for nucleon mass, where  we  look for
 the optimal quark condensate which reproduce 
the nuclon mass within the standard values
 of the condensate $\la \bar{q} q \ra_0(1\rm {GeV}^2)=-(225 \pm 25 MeV)^3$. 

  From ref.\cite{SH95}, we get the FESR for nucleon as follows:
\beq
\label{NFESR}
& & 64 \pi^4 \lambda^2_N= {S_N^3 \over 3}
+ 2 \pi^2 \la {\alpha_s \over \pi} G^2 \ra_0  S_N
+ {128 \over 3} \pi^4 \la \bar{q} q \ra^2_0 
 , \\
& &
64 \pi^4 \lambda^2_N M_N
= -8 \pi^2 \la \bar{q}q \ra  S_\pi^2
 +{32 \over 9} \pi^4 \la \bar{q}q\ra_0 \la {\alpha_s \over \pi} G^2_0 \ra_0 ,\\
& &
\label{NFESRB}
64 \pi^4 \lambda^2_N  M_N^2
= {S_N^4 \over 4}
 +\pi^2 \la {\alpha_s \over \pi} G^2 \ra_0 S_N^2
 -{128 \over 9} \pi^4  \la \bar{q}q\ra^2_0  {\alpha_s  \over \pi} S_N ,
\eeq
where $\lambda_N$ is defined above, and $S_N$ is the continuum threshold of
 nucleon sum rules.  
Solving eq.(\ref{NFESR})-(\ref{NFESRB}) numerically,  we  get results in
 Table 1.
Hence we will utilize  the following numbers in the  anaylses of 
$g_T$ sum rule later; 
\beq
\label{para}
\la \bar{q} q \ra(\GeV^3) =(-0.2185 )^3, \ \ S_N(\GeV^2)=1.6
, \ \ \ \lambda_N ({\rm GeV}^3)=0.0188.
\eeq   
\begin{center}

\begin{tabular}{|c|c|c|c|c|} \hline
$\la \bar{q}q \ra_0$ & $(-0.250\GeV)^3$  & $(-0.230\GeV)^3$  &
 $(-0.2185\GeV)^3$ & $(-0.210\GeV)^3$  \\ \hline
$S_N(\GeV^2)$ & 2.27 & 1.84 & 1.60 & 1.42  \\ \hline
$\lambda_N(\GeV^3)$ & 0.0296 & 0.0224 &  0.0188 &0.0162 \\ \hline
$M_N (\GeV) $ & 1.15 & 1.024 &  0.940  & 0.872\\ \hline
\end{tabular}
\end{center}
Table 1: $S_N, \lambda_N, M_N $ obtained from Eq.(\ref{NFESR}) $\sim$
 (\ref{NFESRB}) with four different values of $\la \bar{q}q\ra_0$ .
\vspace{0.2cm}

Now we turn to  the calculation of 
  induced condensates with the help of
  QSR.
 
We first expand  
 eq.(\ref{cond1}) in terms of   $W^+$ up to first order.  
\beq
 \la \bar{d} \gamma_5 \sigma_{\mu \nu} u \ra_E 
&=&-i {g \over 2 \sqrt2} \int d^4 x \la 0|
 T(\bar{d} \gamma_5 \sigma_{\mu \nu} u (0)
\bar{u} \gamma_\rho \gamma_5 d(x)) |0 \ra \ W^+(x) \\
& =&
\label{cal1-con1}
-i {g \over 2 \sqrt2} \int d^4 x \ \Pai_{\mu \nu, \rho}(x) W^+_\rho(x),
  \nonumber \\
 \mbox{where}& &  \ \ \ \
\Pai_{\mu \nu, \rho}(x) =\la 0| T(\bar{d} \gamma_5 \sigma_{\mu \nu} u(0)
\bar{u} \gamma_\rho \gamma_5 d(x)) |0 \ra. 
\eeq
To estimate $\Pai_{\mu \nu, \rho}(x)$, we expand eq. (\ref{cal1-con1})
 in terms of the local operators 
  up to dimension 5, whose diagrams are Fig.3(a)-(c), and 
 retain the terms proportional to quark mass.   
 Then we get the following equation:
\beq
\label{cal2-con1}
 \Pai_{\mu \nu, \rho} (q) &=
&(-q_\mu g_{\rho \nu}+q_\nu g_{\rho \mu})(m_u-m_d)\chi(q^2) \\
\mbox{with} \ \ &  &   \nonumber \\
& &
\label{con1-ope}
\chi(q^2)={3 \over 8\pi^2} \log(-q^2)
+\left ( {1 \over q^2} + { m_0^2 \over 3 q^4} \right )C_m, 
\eeq
where the first, second and 
  third term on the right hand side  correspond to 
Fig.3(a), (b) and (c), respectively. $m_0^2=0.8({\rm GeV^2})$
 is defined by $\la 0|g_s \bar{q} \sigma \cdot G q |0 \ra
 =m_0^2 \la 0| \bar{q}q|0 \ra $ \cite{BI82}.
Using eq.(\ref{cal1-con1}) and  eq.(\ref{cal2-con1}), 
we reach the result defined in eq.(\ref{cond1}):
\beq
\la \bar{d} \gamma_5 \sigma_{\mu \nu} u \ra_E
={g \over 2 \sqrt2}F_{\mu \nu}(0) (m_u-m_d)\chi(0),
\eeq 
where we replace $F_{\mu \nu}(x) $ by
$ F_{\mu \nu}(0)$ bacause the field strength is assumed to be
 constant.
 For the phenomenological side in eq.(\ref{con1-ope}), 
  we assume that  $\chi(q^2)$ is saturated 
 by 
  $a_1$ meson with mass 1260(MeV),
 which is the lowest state coupled
 to both pseudovector and pseudotensor states,  and  the  continuum
 starting at
$S_\chi$:
\beq
\label{con1-phen}
{1 \over \pi} 
\Im \ \chi(s)= f_\chi \delta(s-m^2_{a_1})-{3 \over 8 \pi^2}\theta(s-S_\chi).
\eeq       
Then, we get 
\beq
\chi(0)={f_\chi \over m_{a_1}^2}- {3 \over 8 \pi^2}
 \int^{\Lambda^2}_0{\theta(s-S_\chi) \over s -m^2_{a_1}} ds,
\eeq
where $\Lambda^2 =1 {\rm GeV^2} $ is taken as a characteristic
 scale of seperating 
 the perturbative and  the non-perturbative part in 
 $\la \bar{d} \gamma_5 \sigma_{\mu \nu} u \ra_E$.
 Matching eq.(\ref{con1-ope}) and eq.(\ref{con1-phen})
  using FESR, we get two sum rules,
\beq
n=0, \ \ \ & &{ 3\over 8\pi^2} S_\chi +C_m=f_\chi, \\
n=1, \ \ \ & &{3 \over 16 \pi^2} S_\chi^2+{m_0^2 \over 3}=f_\chi m^2_{a_1}.
\eeq
 $f_\chi$ is rewritten as
\beq
f_\chi &=& C_m +{3 \over 8\pi^2}m^2_{a_1} \pm 
\sqrt{{C_m \over 4\pi^2} (3 M^2_{a_1}-m_0^2)+
{ 9 m^4_{a_1}\over 64 \pi^4 }}. 
\eeq
To obtain $ f_\chi$, we take $ C_m({\rm GeV}^2)=(-0.0307)-(-0.0223)$ which
 makes the proton-neutron mass difference within the interval $1.95 \GeV \leq
 (M_n-M_p) \leq 2.41 \GeV$
where the electromagnetic effect are subtracted out \cite{JNP95}.
Thus we get    
\beq
\chi(0)=(-0.0337) - (-0.0470)
\eeq

As for $g_s \la \bar{d} \gamma_5 G_{\mu \nu} u \ra_E$ and $
g_s \la \bar{d}\epsilon_{\mu \nu \rho \omega} G^{\rho \omega} u \ra_E$,
  we make OPE up to dimension 7 and get the following results:  
\beq
& & g_s \la \bar{d} \gamma_5 G_{\mu \nu} u \ra_E= {g \over 2 \sqrt2}
C_m(m_u-m_d)F_{\mu \nu} \kappa(0), \\
& & \mbox{with} \ \ \ \kappa(q^2)={m_0^2 \over 12}{1 \over q^2}
+ { \pi^2 \over 36 q^4}\la { \alpha_s \over \pi} G^2 \ra_0, \\
& & g_s \la \bar{d}\epsilon_{\mu \nu \rho \omega} G^{\rho \omega} u \ra_E
=i {g \over 2 \sqrt2}C_m(m_u-m_d)F_{\mu \nu} \xi(0), \\
& &\mbox{with}\ \ \ \xi(q^2)=-{m_0^2 \over 6 q^2}+{1 \over 18 q^4} \pi^2
\la { \alpha_s \over \pi} G^2 \ra_0,
\eeq
where graphs for OPE are shown in Fig.4.
 For phenomenological part,  we adopt two pole approximation namely,
\beq
\kappa(q^2)={f_{a_1} \over m^2_{a_1}- q^2}+{f_{a_2} \over m^2_{a_2}-q^2},
\eeq
where
  $m_{a_1}$ is $a_1$ meson mass 
appeared  above, $m_{a_2}$ is $a_{2}$ meson mass (1360MeV)
 which is the tensor meson, and $f_{a_1}$ and $ f_{a_2}$ represent 
the pole residues.
 The same approximation is also adopted for $\xi(q)$.
  Equating the OPE side  to
the phenomenological side and comparing the coefficients 
up to $ q^4$ to determine
  $f_{a_1}$and $  f_{a_2}$, 
we get $\kappa(0)=-0.079, \ \xi(0)=0.163.$ \\
\section{Analysis and numerical result}
 From  eq.(\ref{ope-final}) and
 (\ref{pheno-final}),  we have found that
 there exist four sum rules corresponding to the tensor structures
  $\hat{p}\sigma_{\mu \nu},\
\sigma_{\mu \nu} \hat{p}, \ \sigma_{\mu \nu},$ and $  
(\gamma_\mu p_\nu-\gamma_\nu p_\mu) \hat{p}$. We
 take only two sum rules among them to evalaute $g_A$ and $ g_T$.

As mentioned above, sum rule for 
$g_A$  is obtained from  the chiral odd structure
 in the chiral limit. On the other hand, sum rule for
  $g_T$  can be deduced from the part proportional to   
$ (\gamma_\mu p_\nu-\gamma_\nu p_\mu) \hat{p}$
 since $ P_4$ in eq.(\ref{pheno-final}) does not contain  $g_A$.
The sum rule obtained from the tensor  structure $\sigma_{\mu \nu}$
 is not suitable, since in  the  phenomenological side
 the contribtion of   the term with $g_A$ is
 comparable to  that of  the term with $g_T$,  and 
 both terms  vanish in the chiral limit.

Matching eq.(\ref{pheno-final}) and eq.(\ref{ope-final}) and making Borel
transform, we obtain  $g_A$ and $g_T$ sum rules in Borel sum rules (BSR)
 \cite{SVZ78} as     
\beq
\left({g_A \over M^2} +A_{sp}\right )& =
&-{e^{{M_N^2 \over  M^2}} \over \lambda_p \lambda_n} \
 \left [{1 \over 8\pi^4}M^4 E_1({S_A \over M^2}) 
-{1 \over 16 \pi^2} \left \la {\alpha_s \over \pi} G^2 \right \ra_0 
-{2 \over 3} {\la \bar{q} q \ra^2_0
\over M^2} \right], \non \\
& &
\label{pre-g_A}
 \\
\left({1 \over M^2} {g_T \over M_p+M_n} +T_{sp}\right )
& =& -{(m_u-m_d) \over \lambda_p \lambda_n} M^2 e^{M_N^2 \over M^2}
\left[ E_0({S_T \over M^2}) \left \{ {1 \over 32 \pi^4}-
{\chi(0) \over 12 \pi^2} \right \}+{C_m \over 8\pi^2 M^2} \right ], \non \\
& &
\label{pre-g_T}
\eeq
where $E_n(x)=1-(1+x+{x^2 \over 2!}+ \cdots + {x^n \over n!})e^{-x}$,
$A_{sp}$ and
 $T_{sp}$ represent the contribution of  single poles coming  from
the nucleon - the resonance transition discussed above.
To subtract $ A_{sp}$ and $ T_{sp}$, 
we multiply the operator ${\partial  \over \partial(1/M^2)}$
 to both sides of eq.(\ref{pre-g_A}) and (\ref{pre-g_T}).
 Assuming that $A_{sp}$ and $B_{sp}$ are
 independent of the  Borel mass $M^2$,  
 we obtain the final sum rules   
\beq
\label{g_A}
g_A& =&{e^{{M_N^2 \over  M^2}} \over \lambda^2_N}{M^6 \over 4 \pi^4}  \\
& \times & \left[E_2({S_A \over M^2})-
 { M_N^2 \over 2 M^2} E_1({S_A \over M^2})-{ \pi^2 M_N^2 \over 4  M^6} 
\left \la {\alpha_s \over \pi}
 G^2 \right \ra_0
+ {8 \pi^4  \la \bar{q}q \ra^2_0 \over 3 M^6}
 \left \{ 1+ {M_N^2 \over M^2}  \right \} \right], 
 \non \\
\label{g_T}
{g_T \over 2 M_N}&=&-{(m_u-m_d) \over \lambda ^2_N} M^2 
e^{M_N^2 \over M^2} \nonumber 
\\
& &\times \left[ \left \{M_N^2
 E_0({S_T \over M^2})-M^2 E_1({S_T \over M^2}) \right \}
\left ( {1 \over 32 \pi^4}-{\chi(0) \over 12 \pi^2} \right )
+{C_m \over 8 \pi^2}{M_N^2 \over M^2} \right], 
\eeq
where $S_A (S_T)$ is the threshold for $g_A (g_T)$, and
 $\lambda_n=\lambda_p \equiv \lambda_N $ and $M_n=M_p \equiv M_N$
    are taken since $m_u-m_d$ is extracted out in (\ref{pre-g_T}).

Here we  make  analyses of  $g_T$ sum rule.
I) We  make an  Borel analysis using nucleon sum rules
 according to the procedures shown in ref.\cite{SVZ78}, and
 apply   the FESR  to get the qualitative understanding.
II) We  make a Borel analysis on the ratio  $g_T/g_A$
 which is
 directly related to the experimental data, using $g_A$ sum rule.
\\
I) First of all, we write down two nucleon sum rules \cite{Ioffe81}
 in order to get rid of the coefficint $e^{M_N^2/M^2}/\lambda^2_N$
   in eq.(\ref{g_T}):
\beq
\label{nucleonQSR1}
4\pi^4 \lambda_N^2 e^{-M_N^2/ M^2}
&=&{M^6 \over 8}E_2(x_N) 
+{\pi^2  M^2 \over 8} \la {\alpha_s \over \pi} G^2 \ra_0 E_0(x_N)
+{8 \pi^4 \over 3} \la \bar{u}u \ra^2 , 
  \\
\label{nucleonQSR2}
4\pi^4 \lambda_N^2 M_N e^{-M_N^2/ M^2}
&=& -\pi^2 \la \bar{d} d \ra  M^4 E_1(x_N) +{2 \pi^4\over 9} 
\la \bar{d} d \ra \la {\alpha_s \over \pi} G^2 \ra_0 
,
\eeq
where $x_N=S_N/ M^2$. We call eq. (\ref{nucleonQSR1})
 (eq.(\ref{nucleonQSR2}))   even (odd) sume rule since
it contains only even (odd) dimensional operators.

 Because the formulas obtained  from BSR  depend on the unphysical
parameter, i.e.,  the Borel mass $M^2$,
we must adopt a Borel window, $M_{min.}^2 < M^2 < M_{max}^2$, in which
$g_T$ is independent of $M^2$ within this range.
 In other words,   
the threshold $S_T$ is chosen to make the Borel curve as flat as possible
 in the Borel window.    
To obtain the Borel window, 
we take $1- E_0(S_T/M^2) \geq 30 \% $ at $S_T=2.0$ 
in eq.(\ref{pre-g_T}) as the upper limit and get $M_{max}^2=1.66$. However,
 we can not get the suitable minimum in  Borel window since
 the contribution of the condensate term in eq.(\ref{g_T})
 is the same magnitude as  that of the perturbative term.    
The multiplication of the operator 
 for subtracting  the single pole makes
 the contribution of the perturbative term 
  reduced because of 
${\partial \over \partial (1/M^2)} M^2 e^{M^2_N / M^2}=0 $ at $M^2=M_N^2$.
Thus, we utilize the lower limit of the nucleon sum rule
 (\ref{nucleonQSR1}) where the second and third terms
  is less
 than 30 \% compared to the pertubative term.
 Thus we arrives at a  Borel window $  1.22 \GeV^2 \leq M^2 < 1.66 \GeV^2$, 
 and we carry out  the Borel analysis on $g_T$ with
 $C_m=$  $-0.0337$ and $ -0.0223$  
to search the optimal threshold
 in the Borel window.

The results are summarized in Table 2 and   Borel curves 
with the optimal threshold are shown in Fig.5 (a) and (d), which 
 show that the magnitude of $g_T$ in our analysis
 is  smaller than
that of the preliminary experimental value by order of magnitude.
To get the physical interpretation of the result,
 we make FESR for  $g_T$ and get the 
 result
\beq
{g_T \over 2 M_N}={(m_u -m_d) \over \lambda^2_N}
\left [ \left ({1 \over 2}S_T^2-M_N^2S_T \right )
\left ({1 \over 32 \pi^4}-{\chi(0) \over 12 \pi^2} \right )
-{C_m \over 8 \pi^2}M_N^2 \right].
\eeq
This implies that adopting $S_T=2 M_N^2 \sim 2.5 M_N^2$ 
as shown in  Table 4
makes the first term with the threshold $S_T$ small compared to
 the second term.
 Neglecting  the first  term and  utilizing
 $\lambda_N^2=4 \la \bar{q} q\ra^2_0$ and $
\ M_N=\left 
({-{25 \pi^2 \over 2} \la \bar{q}q}\ra_0 \right )^{1 \over 3}$
 obtained by FESR,
we arrive at 
\beq
g_T={25 \over 32 }  {\la \bar{d}d-\bar{u}u \ra_0 \over \la \bar{u}u \ra_0 }
={25 \over 32}{C_m (m_u-m_d) \over \la \bar{u}u \ra_0},
\eeq
which gives $g_T=(-0.00896) - (-0.00651)$
 when  $ -0.0307 \leq C_m \leq -0.0223$ is used.
This represents a  good agreement with the result from BSR.

\vspace{0.4cm}
\begin{center}
\begin{tabular}{|c|c|c|c|c|c|} \hline
& $C_m=-0.0307 $  & $C_m=-0.0223 $
& $C_m=-0.0307$  & $ C_m=-0.0223 $
  \\ \hline
     $\la \bar{q}q \ra_0  \ \  \  (S_N )  $
  &  \multicolumn{2}{|c|}{$g_T^{even}\ \ (S_T^{even}) $} &
  \multicolumn{2}{|c|}{$g_T^{odd}\ \ (S_T^{odd})$} 
\\ \hline
$(-0.2185)^3$  (1.60) \ &$-0.0106$ \ (2.15) & $-0.00413$ \ (1.74)  
 & $-0.0163$ \ (2.62) & $ -0.00719 $\ (2.00)  
\\ \hline
\end{tabular}\\
\end{center}
Table 2: $g_T^{even} (g_T^{odd})$  and its threshold 
$S_T^{even} (S_T^{odd})$ using even (odd) nucleon sum rule with
 two different value of $C_m$  where
$\la \bar{q}q \ra_0 $ is in ${\rm GeV^3}$ unit,
 and the thresholds $S_N, S_T$ are in ${\rm GeV^2}$ unit.
 $\la \bar{q}q \ra_0=(-0.2185GeV)^3$ reproduces nucleon mass.\\
\vspace{0.2cm}

II) As it is customary to take the ratio of $g_T$ and $g_A$,
   we  make Borel analysis on $ g_T/ g_A$
  by taking the ratio
 of eq.(\ref{g_A}) and eq.(\ref{g_T}). For the threshold $S_A$,
 we take $S_A$ which satisfies 
 the experimental number of  $g_A$(=1.25) in FESR, and get $S_A=1.68\GeV^2$.
 Then  FESR for $g_A$ reads
\beq
g_A={1 \over 4 \pi^4}{ 1\over \lambda_N^2}
\left[
\frac16 S_A^3-{M_N^2 \over 4} S_A^2-{\pi^2 M_N^2 \over 4} \la {\alpha_s \over \pi} G^2 \ra
+{8 \pi^4  \over 3} \la \bar{q}q \ra^2
\right ] ,
\eeq
where we have used   $\lambda_N$ and $\la \bar{q} q \ra$  in eq. (\ref{para}).
 After searching   optimal threshold in the above  window,  
 we get the results  in Table 3. Corresponding
  Borel curves are given in Fig. 6 where
 we use $\la \bar{q}q \ra_0(\GeV^2) =(-0.2185)^3$
 to compare the results with those in case I).    
Note that replacing  $M^2_{min}=1.22$ by $M^2_{min}=0.61$
 which is obtained from  $g_A$ sum rule
  has the same
  qualitative results with the window used in case  I) 
 and the quantitative change is within 15\%.

 Table 3 shows  that 
\beq
\label{result}
g_T/ g_A = - 0.0152 \pm 0.0053
\eeq
 which   will
 be the one  to be compared with experimental value. 
\vspace{0.4cm}
\begin{center}
\begin{tabular}{|c|c|c|} \hline
 & $C_m=-0.0307$ & $ C_m=-0.0223  $  \\ \hline
$\la \bar{q}q \ra_0 \ \ (S_A)     $
   &  \multicolumn{2}{|c|}{$g_T \ \ (S_T) $} 
\\ \hline
$(-0.2185)^3 (1.68) $ & $-0.0205 \ (2.97)$  &$-0.00983 \ (2.22)$  \\ \hline
\end{tabular}\\
\end{center}
Table 3: $g_T/g_A$  and its threshold $S_{T}$
 with two different values of $C_m$  where
$\la \bar{q}q \ra_0$ is in ${\rm GeV^3}$ unit,
  the thresholds $ S_T, S_A$ are in ${\rm GeV^2}$ unit and 
$C_m$ is in ${\rm GeV^2}$ unit.
 $\la \bar{q}q \ra_0 =(-0.2185 )^3$ reproduces nucleon mass.\\
\vspace{0.3cm}\\
Here we mention the uncertainty of our results originating from $C_m$.
$\chi(0) $  grows as $C_m$ becomes  small, which  changes the
sign of $g_T$ from negative to positive.
Hence determining $g_T$ in QSR  does not become quite accurate  unless 
 $\la \bar{d}d-\bar{u}u \ra_0$ is precisely determined. 
 Also $g_A$  and $g_T$ are    rather sensitive to $\la \bar{q}q \ra_0$, 
 thus we need to know its accurate value.
\\ 
\section{ Discussions and summary}
So far, we have calculated $g_T$  in  QSR and gotten Table 2 and 3, 
and we found that\\
i) $g_T$ is of order  $(m_u-m_d)$.\\
ii) Its  sign based on the definition of
 eq.(\ref{axial-mat}) is negative.\\
 iii)
 $g_T/g_A$ ranges 
from $-0.0205 $ to $ -0.00983 $, i.e., $g_T/g_A \simeq  3 \sim 4 (m_u-m_d)/M_N$
 which is  much smaller than  the preliminary experimental value.\\
iv)
 By using FESR, we get the analytic formula for $g_T$:
\beq
g_T={25 \over 32 }  {\la \bar{d}d-\bar{u}u \ra_0 \over \la \bar{u}u \ra_0 }
={25 \over 32}{C_m (m_u-m_d) \over \la \bar{u}u \ra_0}.
\eeq

As mentioned  in the introduction, 
  the MIT bag model  has been  utilized so far to  calculate $g_T$. 
In this model, we get the following result up to
 $O( m_u-m_d)$ ( see Appendix A for the detailed calculations): 
\beq
g_T=0.041 M_N (m_u -m_d)R^2, 
\eeq
 with $R$ being the bag radius. Note that the difference of the
 bag radius  between the proton and the neutron is neglected.
By taking $R=1.085$ fm \cite{DGH92},  which reproduces the 
proton mass, 
 we obtain $g_T=-0.00455$ which is consistent with
 the result obtained by QSR.
Since the obtained result is  rather  sensitive to the bag radius, 
one should take this number only qualitatively.
   
Another  effect to  $g_T$, which we must take into account, is 
the  electromagnetic effect.    
Rough estimate  using a hadronic model 
shows that this  effect is smaller
 than that of the u-d quark mass difference   
as in the case of the $\rho-\omega$ mixing and the p-n mass difference.
Thus our conclusion eq.(\ref{result}) will not be changed  qualitatively
by the electromagnetic effect.
Nevertheless, more detailed study of QSR with the
electromagnetic effect must be done. 

In summary we have examined the induced tensor, $g_T$, 
 in QSR with the external field, and gotten
 $g_T/g_A=$-$0.0152 \pm 0.0053$ which is smaller than 
  preliminary  experimntal numbers by one order of magnitude.  
(current experimental number is ranging from $0.14 \pm 0.10$
 to $-0.21 \pm 0.14$ \cite{Morita9}).
However,  the experiments and its analyses
  remain uncertain in order  to compare with  
    result obtained in this paper
 \cite{Morita9,Hol89,nwp}.
Our result should be    checked in future beta-decay experiments
 to understand the  G-parity violation.    
\acknowledgments
I would like to thank  T. Hatsuda for his useful and valuable discussions
 throughout this work.  
Also, I would like to thank  M. Doi and  M. Morita
 for drawing my attention to the induced tensor.  

\newpage
\appendix
\setcounter{equation}{0}
\section{Estimate of Induced tensor in the MIT bag model}
We define the matrix element as
\beq
\la p(p_2)| J^5_\mu(x) |n(p_1) \ra = 
\bar{u}_p(p_2)
 \left ({i \gamma_5 \sigma_{\mu \nu} \over M_n+M_p}g_T q_\nu \right)
u_n(p_1)
\eeq
where  $q=p_1-p_2$.\\
I)
\beq
\label{Bag1}
 {\partial \over \partial \vec{q}} \la p| J^5_0 (x) |n \ra|_{\vec{q}=0}
={g_T \over 2M_N} \bar{u}_p \vec{\sigma} u_n
={g_T \over 2M_N} {}_{s-f}\la p| \vec{\sigma} \tau^+|n \ra_{s-f},
\eeq
where the subscript denotes a spin-flavor matrix element and, we use
\beq
 \la p| J^5_\mu(x) |n \ra={g_T \over 2 M_N} \vec{q} \bar{u}_p
\left(
\begin{array}{cc}
\vec{\sigma} & 0 \\
0 & \vec{\sigma}
\end{array} \right) u_n \nonumber 
\eeq
II)
The axial current in the MIT bag model is
\beq
J^5_\mu(x)=  \sum_i \bar{q}_i(x) \gamma_\mu \gamma_5 \tau^+ q_i(x)
\eeq
From eq.(\ref{Bag1}), we obtain
\beq
 {\partial \over \partial \vec{q}} \la p| J^5_0(x) |n \ra |_{\vec{q}}
&=&{\partial \over \partial \vec{q}}
 \int \ d^3 x \ \la p| J^5_0(x)|n \ra e^{-i \vec{q} \cdot x} |_{\vec{q}=0}
 \non \\
&= &  -i \int \ d^3 x \ \vec{x}
\la p| \sum_i \bar{q}_i(x)  \gamma^0 \gamma_5 \tau^+ q_i(x) |n \ra.
\eeq
Hence,
\beq
& &
\label{Bag2}
\int_{bag} d^3 x \bar{q}_u(x) \gamma^0 \gamma^5 q_d(x) \vec{x}  \\
&=&
{N_u N_d \over 4 \pi} \int_{bag} d^3 x \hat{x} \
\left \{ \sqrt{E_u+m_u \over E_u} \sqrt{E_d-m_d \over E_d}
 j_0(x_u r/R) i \vec{\sigma} \cdot \hat{r} j_1(x_d r / R) 
-(u \leftrightarrow  d) \right \} \non
\eeq
where
\beq
& &q(x)={1 \over \sqrt{4 \pi}} \left[
\begin{array}{c}
\sqrt{E+m \over E} j_0(xr/R) \non \\
\sqrt{E-m \over E} i \vec{\sigma} \hat{r} j_1(xr/R)
\end{array} \right]
, \ \ \
E(m, R)={1 \over R} [x^2 +(mR)^2]^{1/2} \\
& & N^{-2}_q (x)=R^3 j_0^2(x) {2 E(E- 1/R) +m/R \over E(E-m)}, \ \ \
\tan x={x \over 1 -mR-[x^2 +(mR)^2]^{1/2}}, \non 
\eeq
where $j_n(x)$ is spherical Bessel function and $R$ is Bag radius
and $r=|\vec{x}|$.\\
To show the effect of $u-d$ quark mass difference,
 we expand eq.(\ref{Bag2}) up to linear in $m_u-m_d$:
\beq
\mbox{(Eq.\ref{Bag2})}&= &{(m_u-m_d)R \over x_0-1}{N^2 \over 4\pi}
\int_{bag} \  d^3 x \ r j_0(x_0 r/R) j_1(x_0 r/R) \non \\
 &=&
-{(m_u-m_d)  \over 2(x_0-1)} {N^2 \over 4 \pi}
\int_{bag}\ d^3 x \ r^2 (j_0^2 (x_0 r/R)+j_1^2(x_0 r/R) ) 
\eeq
where 
\beq
\tan x_0={x_0 \over 1-x_0}, 
\ \ \
E(m, R)={x_0 \over R}+{m \over 2(x_0-1)}+O(m^2).
\eeq
Thus, we get
\beq
 & &{\partial \over \partial \vec{q}} \la p| J^5_0(x) |n \ra |_{\vec{q}}
 \non \\
&\times&
{1 \over 3} {}_{s-f}\la p| \sum_i \vec{\sigma}_i \cdot \tau^+_i |n \ra_{s-f}
{(m_u-m_d)N^2 \over 4 \pi(x_0-1)} \non \\
&=& 
\label{Bag3}
\left( R \int d^3 x \ r j_0(x_0 r/R) j_1(x_0 r/R)
   -{1 \over 2 } \int d^3 x \ r^2 \ (j_0^2(x_0 r/R)+j_1^2(x_0 r/R) \right).
\eeq  
Equating eq.(\ref{Bag1}) to eq.(\ref{Bag3}), we arrive at
\beq
{g_T \over 2 M_N}=-(m_u-m_d){5 R^2 \over 36 x_0 (x_0^2-1)^2}
\left [ {1 \over x_0}-{17 \over 6}+{8 \over 3}x_0 -{2 \over 3} x_0^2 \right],
\eeq
where $x_0=2.04 $ and we have used
\beq 
{}_{s-f}\la p| \sum_i \vec{\sigma}_i  \cdot \tau^+_i |n \ra_{s-f}
={5 \over 3}{}_{s-f}\la p|  \vec{\sigma} \cdot \tau^+ |n \ra_{s-f}.
\eeq
\newpage
\centerline{Figure Captions}
\noindent
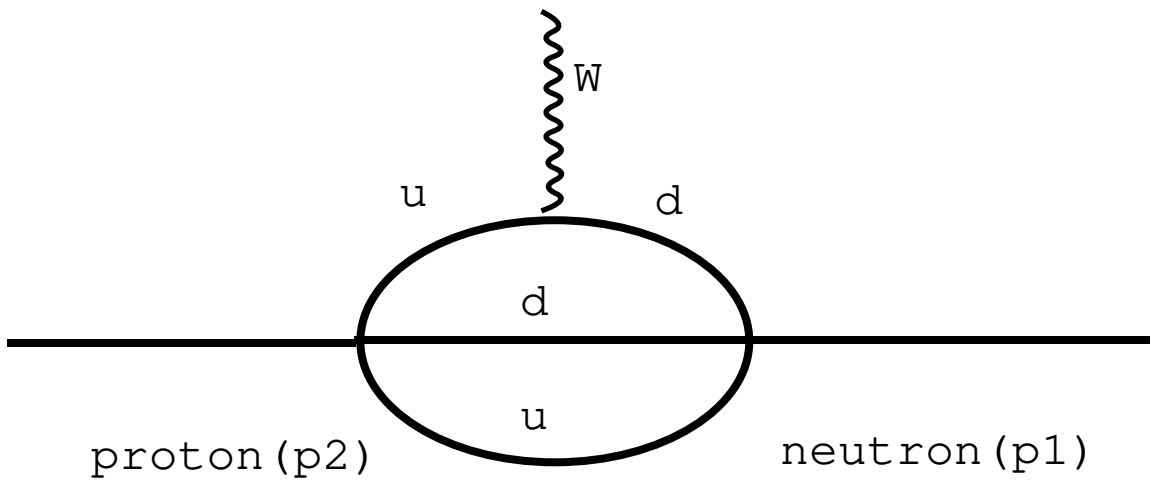
{\bf Fig.1}\\
A schematic illustration  that neuton absorbes $W^+$ boson and
turns into proton.\\
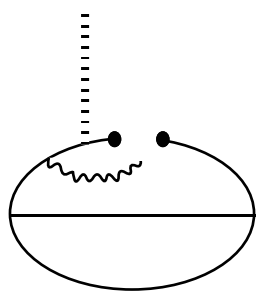
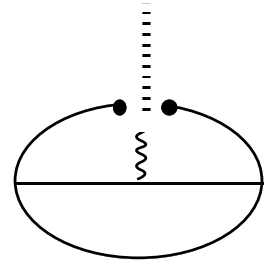
{\bf Fig.2}\\
OPE for
 $\Pai_E(p)$, where for the chiral odd structures
   $\Pai_E(p)$ is expanded
 up to dimension 8 with $m_u=m_d$, while for the chiral even strutures
$\Pai_E(p)$ is expanded up to dimension 5, and up  to linear in
$(m_u-m_d)$.
 Dashed lines denotes the external field,
 wavy lines denote gluon lines,and broken lines denote
 the quark/gluon condensate.\\
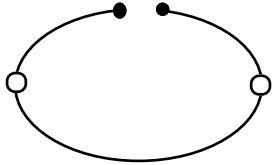
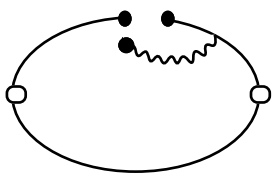
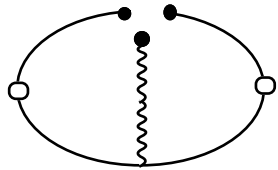
{\bf Fig.3}\\
OPE  up to dimension 5 for $\chi(p)$ sum rules.
 Wavy lines denote gluon lines and broken lines denote
 the quark/gluon condensate.\\
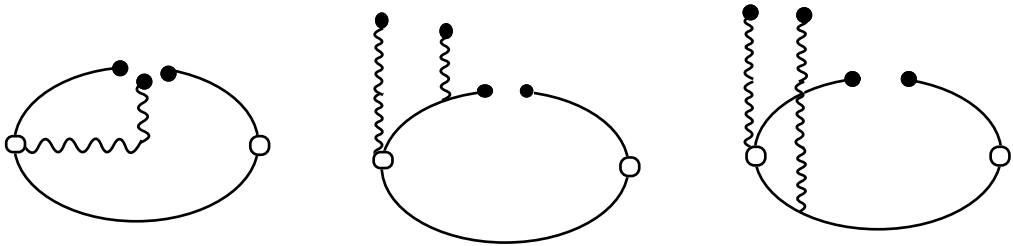
{\bf Fig.4}\\
OPE  up to dimension 7 for $\kappa(p)$ and $ \xi(p)$  sum rules.
 Wavy lines denote gluon lines and  broken lines denote
 the quark/gluon condensate.\\
{\bf Fig.5 (a),(b)}\\
 $ g_T^{even}$ and $g_T^{odd}$ with the optimal threshold $S_T$
as a function of the Borel mass squared $M^2$.
$S_T$ is also shown in ${\rm GeV}^2$ unit.
(a)((b)) corresponds to $g_T^{even} (g_T^{odd}) $. 
 The solid ( dashed) line  corresponds
 to  $C_m(\GeV^2)=-0.0306$ ( $-0.0223 )$,
\\
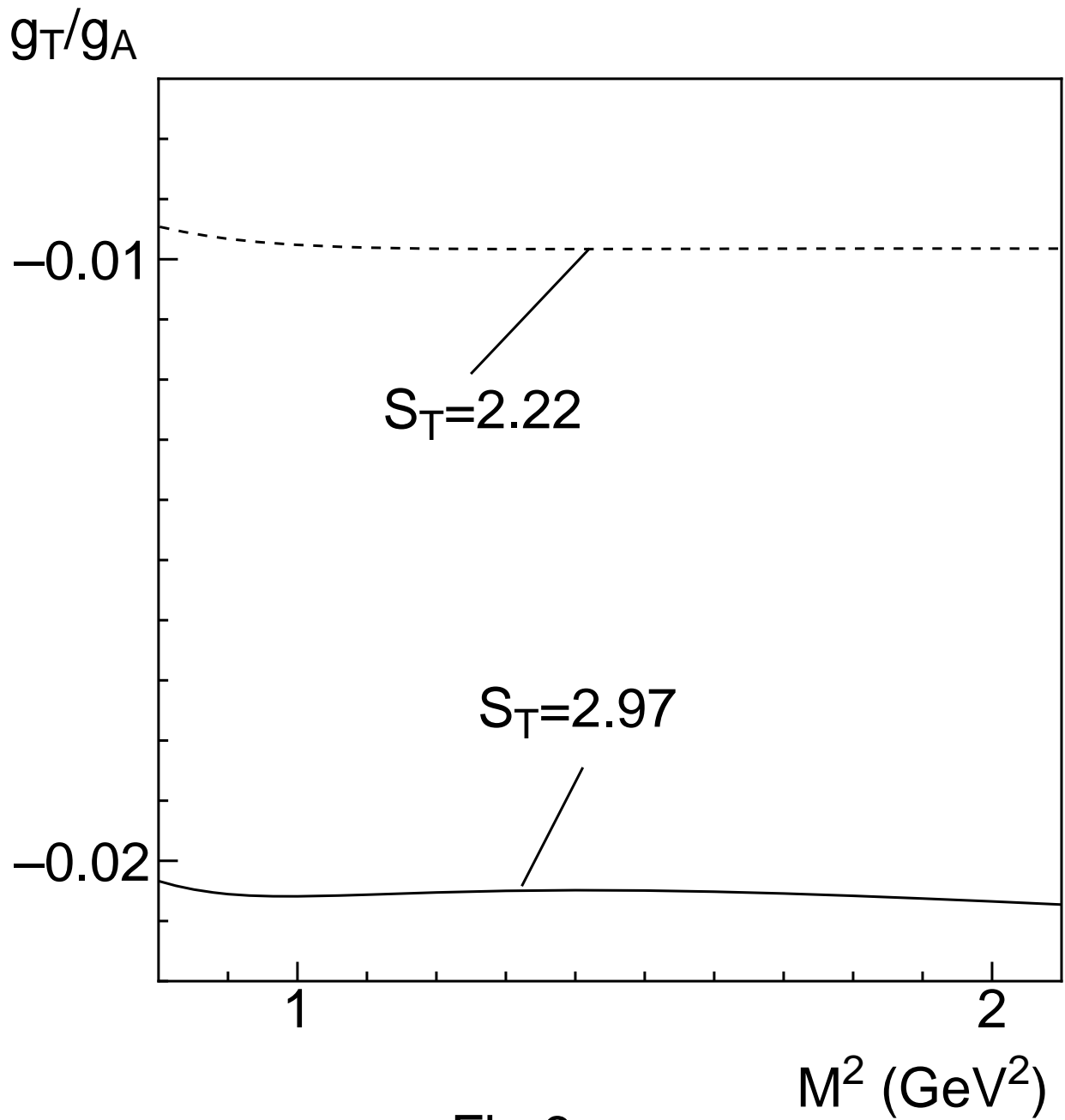
{\bf  Fig.6}\\
 $ g_T/g_A$ with the optimal threshold $S_{T}$
as a function of the Borel mass squared $M^2$.
$S_{T}$ is also shown in ${\rm GeV}^2$  unit.
 The solid ( dashed) line  corresponds
 to  $C_m(\GeV^2)=-0.0306 (-0.0223) $.
\newpage
%

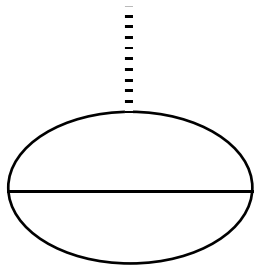
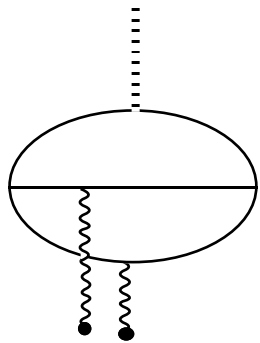
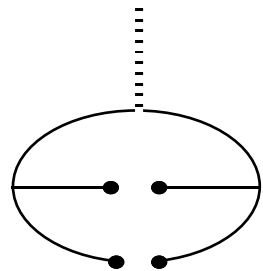
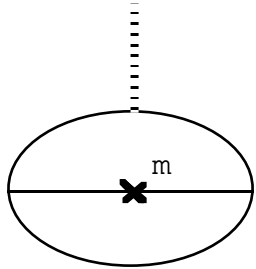
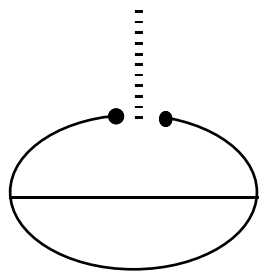
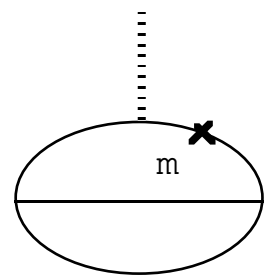
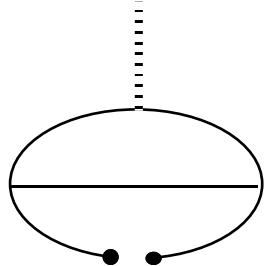
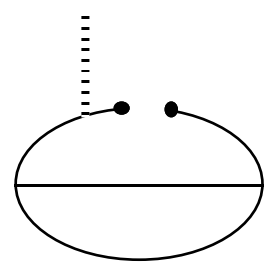
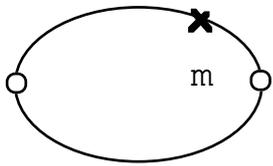
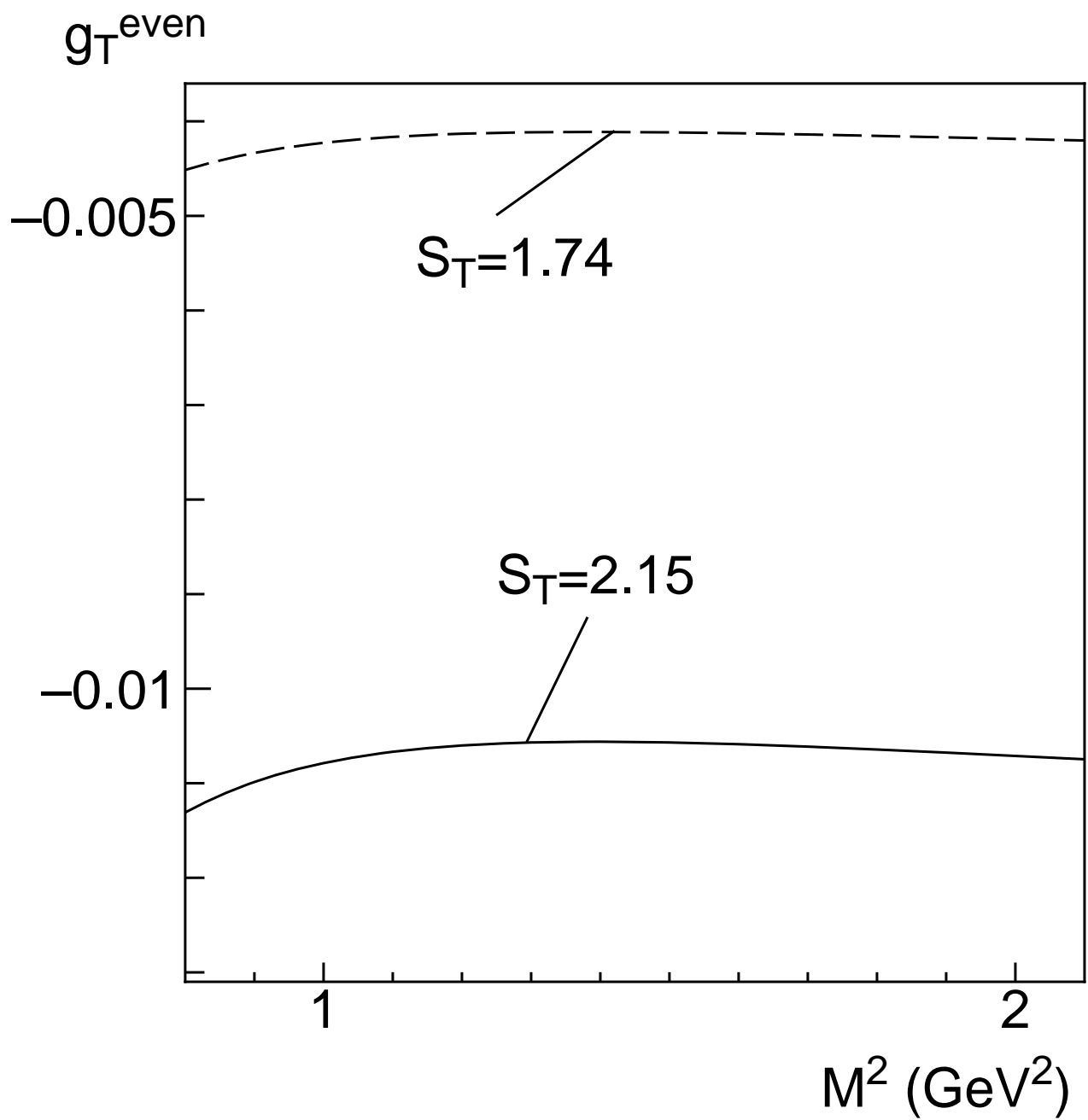
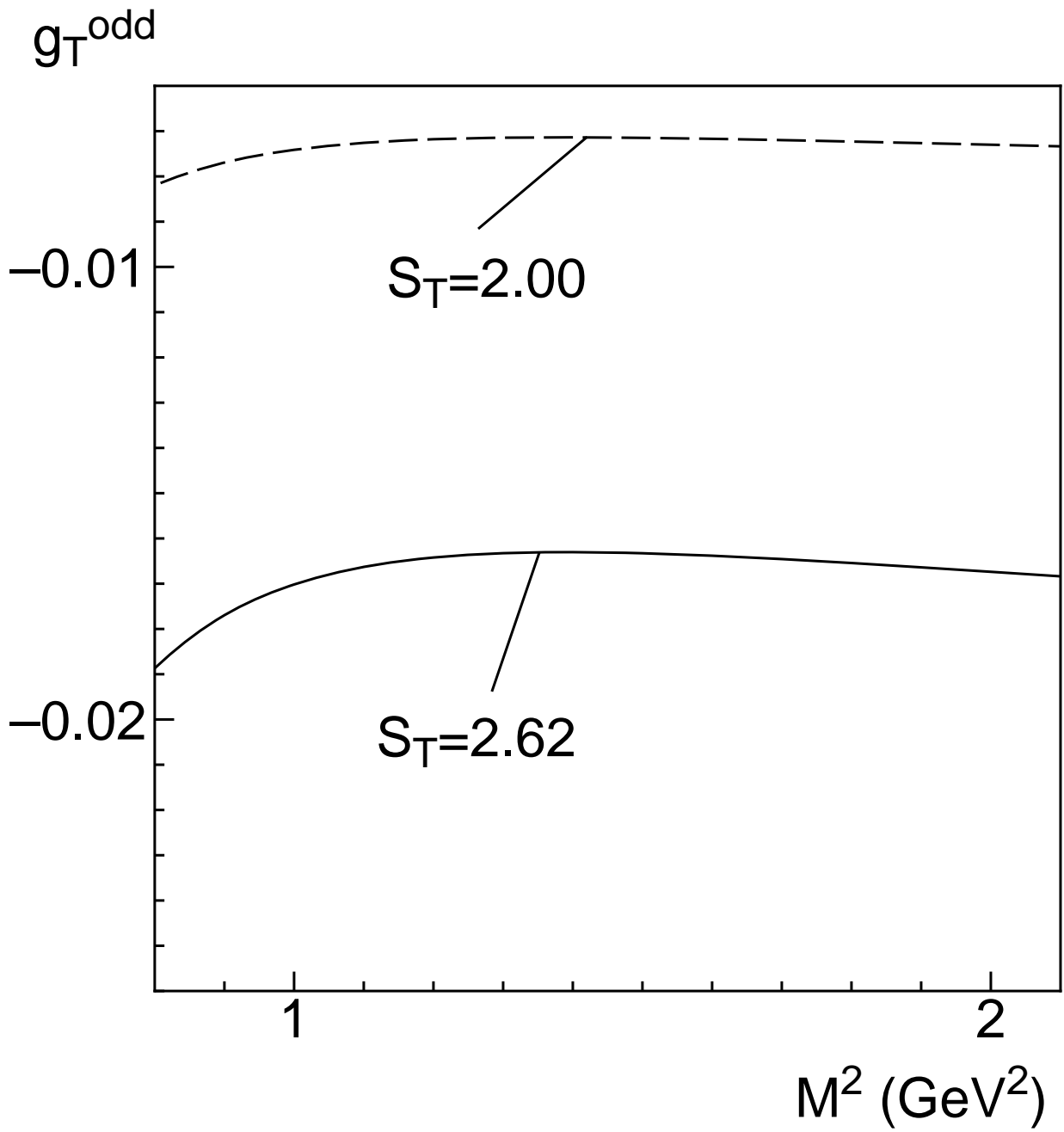
\end{document}